# Two-dimensional Dirac signature of germanene

L. Zhang, P. Bampoulis, A. van Houselt and H.J.W. Zandvliet

Physics of Interfaces and Nanomaterials group, MESA+ Institute for Nanotechnology and University of Twente, P.O. Box 217, 7500AE Enschede, The Netherlands

The structural and electronic properties of germanene coated $Ge_2Pt$ clusters have been determined by scanning tunneling microscopy and spectroscopy at room temperature. The interior of the germanene sheet exhibits a buckled honeycomb structure with a lattice constant of 4.3 Å and a buckling of 0.2 Å. The zigzag edges of germanene are reconstructed and display a 4× periodicity. The differential conductivity of the interior of the germanene sheet has a V-shape, which is reminiscent of the density of states of a two-dimensional Dirac system. The minimum of the differential conductivity is located close to the Fermi level and has a non-zero value, which we ascribe to the metallic character of the underlying $Ge_2Pt$ substrate. Near the reconstructed germanene zigzag edges the shape of the differential conductivity changes from a V-shape to a more parabolic-like shape, revealing that the reconstructed germanene zigzag edges do not exhibit a pronounced metallic edge state.

In the past decade a new class of materials has been developed, which is not three-dimensional (3D), but two-dimensional (2D) in nature. Graphene is by far the most famous example of this new class of 2D materials [1,2]. Graphene consists of a single layer of *sp²* hybridized carbon atoms that are arranged in a planar honeycomb registry. Graphene is a very appealing material because of its unique physical properties [1,2]. The charge carriers in graphene behave as relativistic massless particles that are described by the Dirac equation, i.e. the relativistic variant of the Schrödinger equation. In the vicinity of the Dirac point the dispersion relation is linear, i.e. $E = v_F \eta k$, where $v_F$ is the Fermi velocity, $\eta$ the reduced Planck constant and *k* the wave vector. Graphene is a semimetal and the density of states scales linearly with energy. One of the interesting properties of finite graphene sheets is the existence of electronic states that are localized at the edges of graphene. Theory predicts that a zigzag terminated graphene edge is metallic, whereas an armchair terminated graphene edge is semiconducting [3-5]. Scanning tunneling microscopy and spectroscopy studies of zigzag and armchair monatomic step edges of graphite have indeed confirmed these theoretical predictions [6-8].

Since the rise of graphene there has been a growing interest in other two-dimensional materials that exhibit 'graphene'-like properties. The most obvious alternatives for graphene are the group IV elements, i.e. silicon, germanium and tin. Unfortunately, these graphene analogues of silicon (silicene), germanium (germanene) and tin (stanene) do not occur in nature and therefore these materials have to be synthesized. Germanene is one of the youngest members of the graphene family and has not been studied extensively. In contrast to the planar graphene lattice, the germanene honeycomb lattice is buckled. Theoretical calculations have shown that despite this buckling the 2D Dirac properties of germanene are preserved [9]. Only just recently a few research groups have managed to synthesize germanene. Li et al. have grown germanene on Pt(111) substrates [10], Dávila et al. on Au(111) substrates [11], Bampoulis et al. on $Ge_2Pt$ nanocrystals [12] and finally Derivaz et al. on Al(111) substrates [13]. In particular the work of Dávila et al. [12] provides a detailed study of the growth of germanene involving a large arsenal of experimental techniques and density functional theory calculations. To date the focus of the experimental work has been on the synthesis and the structural properties of germanene, whereas the electronic properties of germanene remained almost completely unexplored.

Here we will study the structural and electronic properties of germanene sheets that are found on $Ge_2Pt$ crystals after deposition of Pt on Ge(110) substrates. We will show that the density of states of these germanene sheets hints to a 2D Dirac system, albeit the density of states does not completely vanish at the Fermi level. We will also study the structural properties as well as the density of states of the edges of the germanene sheets. We found zigzag edges of germanene that exhibit a *4a* periodicity. Surprisingly, we did not find any evidence for the existence of a metallic edge state.

The experiments have been performed in an ultra-high vacuum system with a base pressure of $3 \times 10^{-11}$ mbar. The system is equipped with room temperature scanning tunneling microscope purchased from Omicron (STM-1). The lightly doped *n*-type Ge(110) samples were cut from nominally flat 10x10x0.4 mm, single-side-polished substrates. The Ge(110) substrates were mounted on Mo sample holders and contact of the samples to any other metal during preparation and experiment has been carefully avoided. After cleaning and mounting of the samples, they have been outgassed in ultra-high vacuum at 700 K for about 24 hours. Subsequently, the Ge(110) substrates were cleaned by a cleaning method that has been successfully applied to the more abundant Ge(001) and Ge(111) crystals [14]. This method involves several cycles of Argon ion sputtering at 500-800 eV and annealing at 1100 (±25) K. After checking the cleanliness and flatness of the substrate with scanning tunneling microscopy Pt was deposited onto the substrate at room temperature. Pt was evaporated by resistively heating a W wire wrapped with high purity Pt (99.995%). After Pt-deposition the sample was shortly annealed at 1100 (±25) K and subsequently cooled slowly to room temperature before placing it into the scanning tunneling microscope for imaging. Tapping mode atomic force microscopy (AFM) was performed using an Agilent 5100 atomic force microscope (Agilent) and HI'RES-C14/CR-AU probes (MikroMasch), with a nominal spring-constant of 5 N/m and resonance frequency of 160 kHz.

A few monolayers of Pt were deposited on a Ge(110) substrate at room temperature and subsequently the substrate was annealed at a temperature of about 1100 K for 10 seconds. In a previous paper [12] we have described that at temperatures higher than ~1000 K eutectic $Pt_{0.22}Ge_{0.78}$ droplets are formed on the Ge(110) substrates. Upon cooling down, the eutectic droplets undergo spinodal decomposition into a pure Ge phase and a $Ge_2Pt$ phase The Ge phase segregates to the surface, whereas the $Ge_2Pt$ remains at the interior of the solidified droplets. After slowly cooling down the sample to room temperature it was imaged with an atomic force microscope. We found several types of $Ge_2Pt$ crystals: pyramidal and flat-topped shaped crystals. In Figure 1 large scale images of a pyramidal shaped and a flat-topped $Ge_2Pt$ crystals are shown. The elongated $Ge_2Pt$ crystal has a length of 1.6 µm, a width of 0.5 µm and a height of 80 nm. The flat-topped $Ge_2Pt$ crystals are terminated by a buckled honeycomb lattice with a lattice constant of 4.3±0.1 Å (see Figure 2A for a small scale scanning tunneling microscopy image). The buckled honeycomb lattice is composed of two triangular sub-lattices that are displaced with respect to each other in a vertical direction by ~ 0.2 Å. This buckling is significantly smaller than the 0.6-0.7 Å predicted for germanene by Cahangirov et al. [9] and Garica et al. [15] using density functional theory calculations. However the nearest-neighbor distance between the atoms of the honeycomb lattice is 2.5±0.1 Å, i.e. very close to the predictions for free-standing germanene [9,15].

Scanning tunneling spectroscopy was performed on a 13 nm × 13 nm area. At every point of a 60 by 60 grid an *I(V)* curve was recorded with the feedback loop disabled. The voltage was swept between -1 V and 1V. The set point tunnel current at -1 V was 0.6 nA. The differential conductivity, *dI/dV*, was obtained by numerically differentiating the *I(V)* curves. Subsequently, we took the average of all 3600 *dI/dV* curves (see Figure 2B). The *dI/dV* curve, which is proportional to the local density of states, has a well-defined V-shape reminiscent of the density of states of a 2D Dirac system. The slope of the *dI/dV* curve is, however, a bit more steeper for the unoccupied states than for the occupied states. At first glance this seems at variance with a 2D Dirac system, for which the density of states should be fully symmetric around the Dirac point due to electron-hole symmetry. Two remarks regarding the shape of our *dI/dV* measurements are in place here. First, our *dI/dV* spectra are not recorded on a free-standing germanene sheet, but on a germanene sheet on a metallic $Ge_2Pt$ support. Second, also the electronic structure of the W tip has an effect on the measured *dI/dV*. The observed asymmetry of the *dI/dV* curves can therefore also be due to the electronic structure of the underlying $Ge_2Pt$ substrate and/or W tip. We have investigated several samples and used various STM tips and in virtually all cases we find slightly asymmetric *dI/dV* curves. In some cases the density of states at energies below the Fermi level was higher than the density of states above the Fermi level, whereas in other cases we observed a reversed asymmetry (see Figure 2C). However, in all cases the spatial variation of the *dI/dV* curves within a scanning tunneling spectroscopy grid scan is very uniform. The non-zero value of *dI/dV* at the minimum of the V-shaped curve is ascribed to the underlying metallic $Ge_2Pt$ substrate. The energy range of the V-shaped density of states is about 1 eV, which is significantly larger than the predicted 0.5-0.8 eV energy range of the Dirac cone for free-standing germanene [9]. As a possible explanation for this large energy range we suggest that the electronic states near the $\Gamma$ point of the germanene Brillouin zone are affected by the underlying $Ge_2Pt$ substrate. As a final remark regarding the shape of the *dI/dV* spectrum we would like to emphasize that near the minimum of *dI/dV* curve the shape deviates from a perfect V-shape due to thermal broadening.

Additional support for the linear dispersion relation of our germanene sheet can, in principle, be obtained from an angle-resolved photoemission study. Unfortunately the small size of our $Ge_2Pt$ crystals (<1-2 μm) does not allow us to perform angle-resolved photoemission experiments. This is a pity because these experiments would also allow us to determine the Fermi velocity from the slope of the dispersion curve.

In Figure 3A a scanning tunneling microscopy image of germanene sheet is shown. Line scans taken across and along the step edge are depicted in Figures 3B and 3C, respectively. The monatomic step height is 0.56 nm and agrees well with observations by Bampoulis et al. [12]. A line scan taken along the step edge reveals a 4*a* periodicity, where *a*=0.43 nm is the lattice constant of germanene. The edges of germanene can be divided into two types of edges, the so-called zigzag and armchair edges. Although we did not succeed to obtain atomic resolution at the step edges, we determined the orientation of the step edges from atomically resolved images recorded on the flat terraces. Based on our observations we conclude that the step edge in Figure 3A is a zigzag type of step edge. The unreconstructed zigzag edge consists of a chain of hexagons. However, the zigzag edge of graphene can lower its energy by forming a chain of pentagon-heptagon pairs, the so-called Stone-Wales defects [16-18]. This reconstruction of the zigzag edge is referred to as the zz(57) [16] or the Stone-Wales reconstruction. The zz(57) reconstruction leads to a doubling of the periodicity of the edge. The periodicity of the germanene zigzag edge in Figure 3C has, however, a periodicity of 4*a*. Since our resolution is insufficient to resolve the exact details of the zigzag edge we cannot determine the exact structure of the zigzag edge. However, it is not difficult to come up with simple structural models that are for instance composed of alternating units of zz(57) and zz(66) building blocks (see Figure 4 for a model of the zz(5766) edge). As a final remark we would like to mention that besides the abundant reconstructed zigzag edges we also found several rough edges (see the inset of Figure 3A for an example).

Near the step edge the differential conductivity is smaller than at the terraces (see Figure 3D). The shape of the *dI/dV* curve has also changed substantially from an asymmetric V-shape to a flatter, more parabolic, shape. Interestingly, our observations are very similar to results obtained by Liu et al. [19] for graphene sheets on a hexagonal boron nitride layer. In any case the reconstructed germanene zigzag edges do not exhibit a pronounced metallic edge state.

As pointed out by Koskinen, Malola, and Häkkinen [16] there are two types of metallic zigzag edges states. The zigzag edge of graphene has a metallic edge state that does not stem from the dangling bonds at the edge, but from the topology of the π-electron network. The energy band of the metallic edge state of the zigzag edge is rather flat resulting in a well-defined peak in the density of states at the Fermi level. The armchair edge does not possess such a metallic edge state. The other metallic edge state comes from the dangling bonds of the edge atoms.

As a final remark we would like to stress that the $Ge_2Pt$ substrate is not an ideal substrate for germanene because of its metallic character. A wide band gap material, such as hexagonal boron nitride (h-BN) or aluminum nitride (AlN), is a much more appealing substrate for germanene. These substrates allow to decouple the important electronic states of the germanene near the Fermi level from the underlying substrate. h-BN is probably the best candidate because the lattice constant of the h-BN lattice is very comparable to the nearest-neighbor distance of the germanene lattice.

The structural and electronic properties of germanene sheets synthesized on $Ge_2Pt$ crystals are studied at room temperature with scanning tunneling microscopy and spectroscopy. The interior of the germanene sheets displays a V-shaped density of states indicative for a 2D Dirac system. However, the density of states at the minimum of the V-shaped curve does not completely vanish, which we ascribe to the underlying metallic $Ge_2Pt$ substrate. Two types of step edges are found on the germanene sheets: straight reconstructed zigzag step edges with a 4× periodicity and rough step edges. Both type of step edges have a parabolic shaped *dI/dV* curve.

**Acknowledgments**

Lijie Zhang thanks the China Scholarship Council for financial support. Pantelis Bampoulis thanks the Nederlandse organisatie voor wetenschappelijk onderzoek (NWO) for financial support. Harold Zandvliet thanks the Stichting voor Fundamenteel Onderzoek der Materie (FOM) for financial support.

**Figure captions**

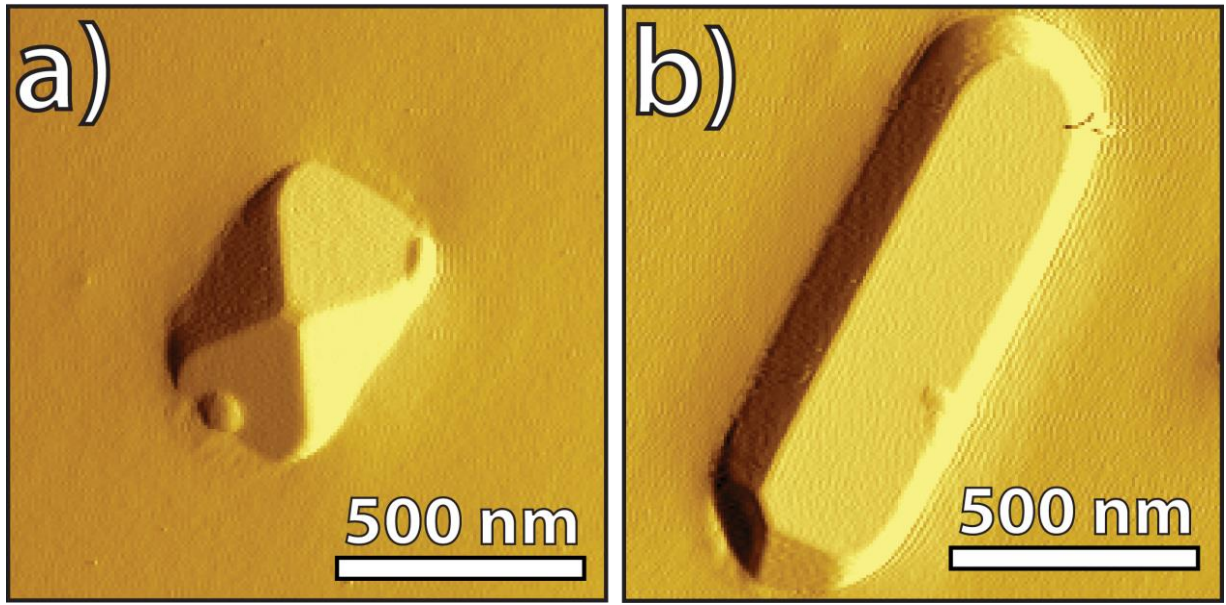

**Figure 1**
(a) Large scale atomic force microscopy image of a pyramidal shaped $Ge_2Pt$ crystal. The height of the cluster is 105 nm.
(b) Large scale atomic force microscopy image of a flat-topped $Ge_2Pt$ crystal. The height of the cluster is 80 nm.

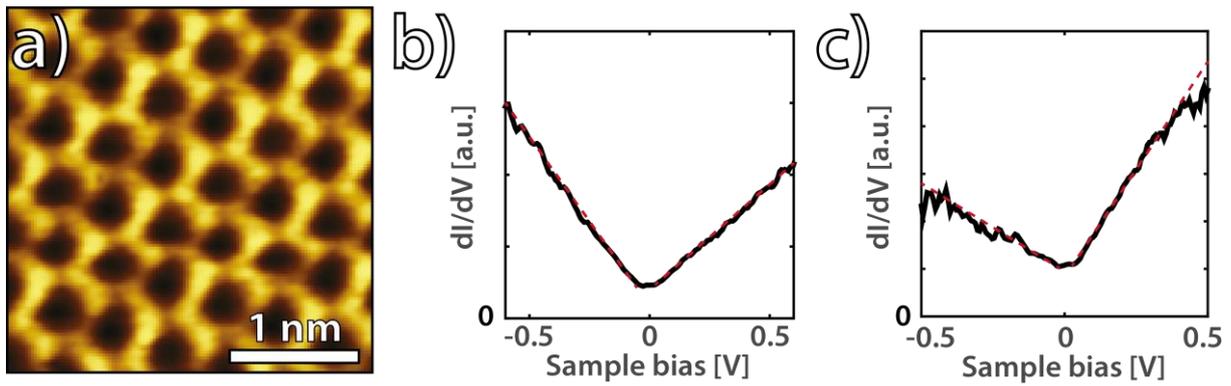

**Figure 2**

(a) Small scale (3 nm × 3 nm) scanning tunneling microscopy image taken at the flat top plane of one of the germanene terminated $Ge_2Pt$ crystals. Sample bias -0.5 V, tunnel current 0.2 nA.

(b) Differential conductivity recorded at the interior of a germanene sheet. The dI/dV curve is averaged over a 13 nm by 13 nm area using a 60 × 60 grid. Set point voltage is -1 V and set point tunnel current is 0.6 nA.

(c) Differential conductivity recorded at the interior of a germanene sheet. The dI/dV curve is averaged over a 13 nm by 13 nm area using a 60 × 60 grid. Set point voltage is -0.54 V and set point tunnel current is 0.52 nA.

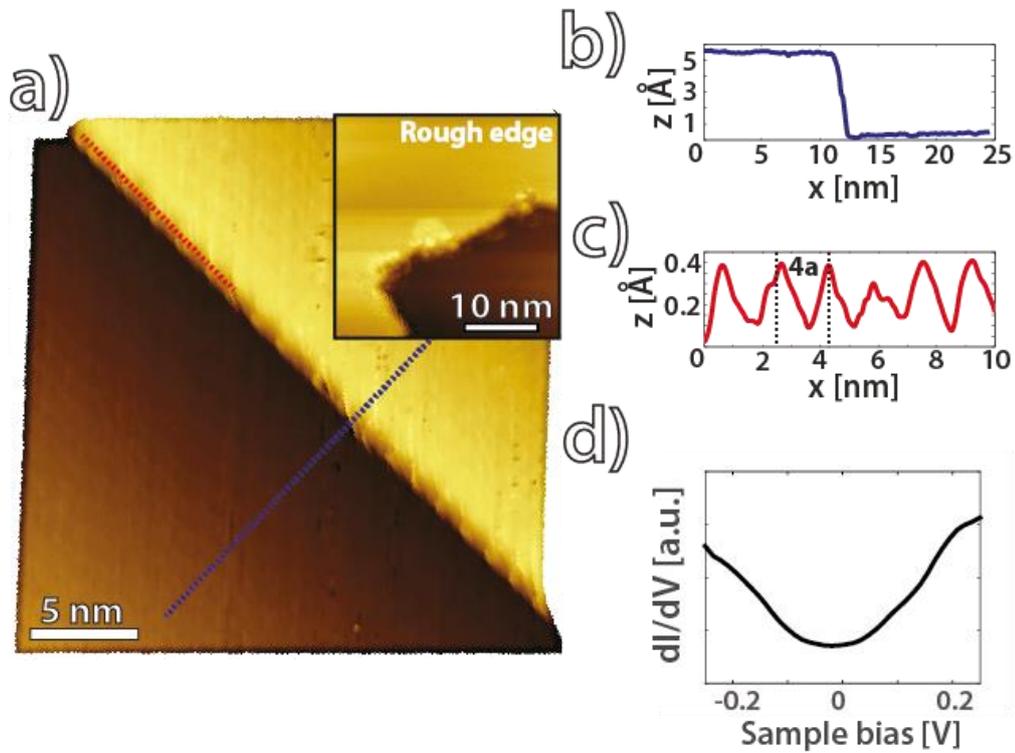

**Figure 3**

(a) Scanning tunneling microscopy image of a germanene step edge (25 nm × 25 nm). Inset: scanning tunneling microscopy images of a rough edge.

(b) Line scan taken across the step edge (blue dotted line). The monatomic step height is 0.56 nm.

(c) Line scan taken along the step edge (red dotted line). The periodicity is 4$a$, where $a$=0.43 nm is the lattice constant of germanene

(d) Differential conductivity recorded near the step edge. Set point voltage -1.57 V and set point tunnel current 0.47 nA.

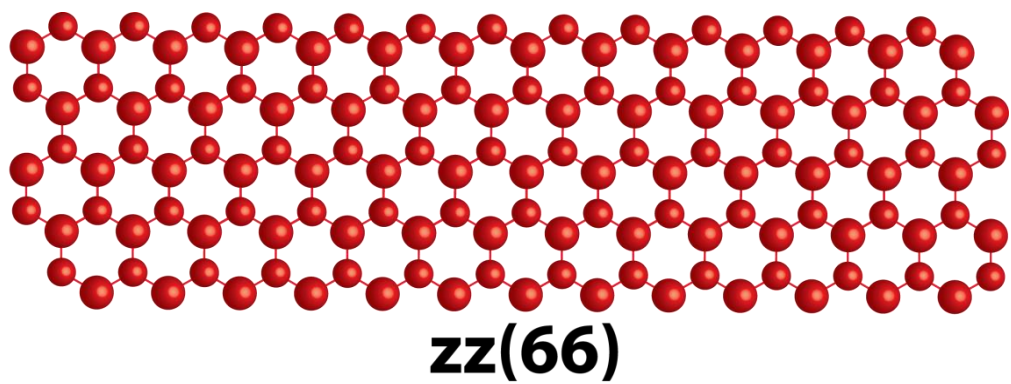

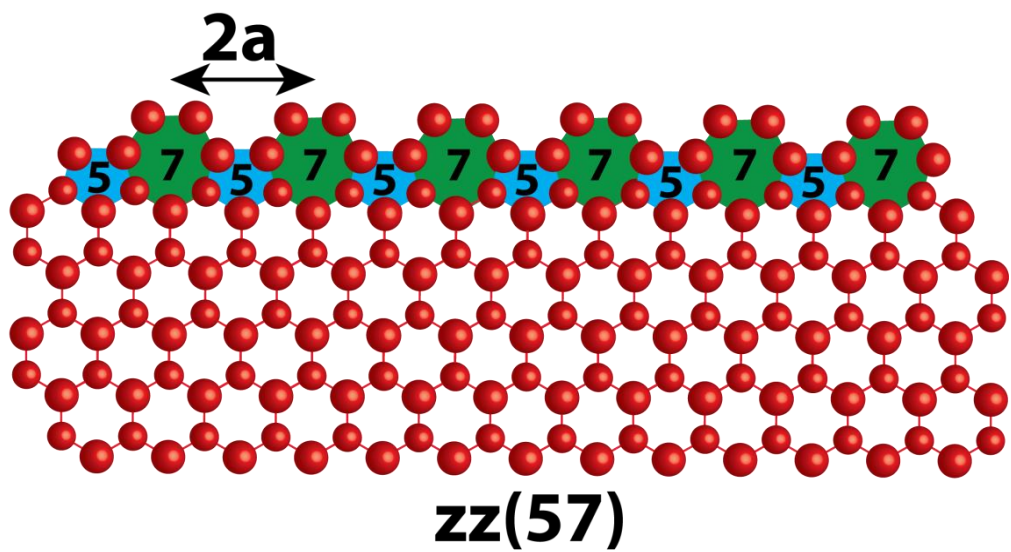

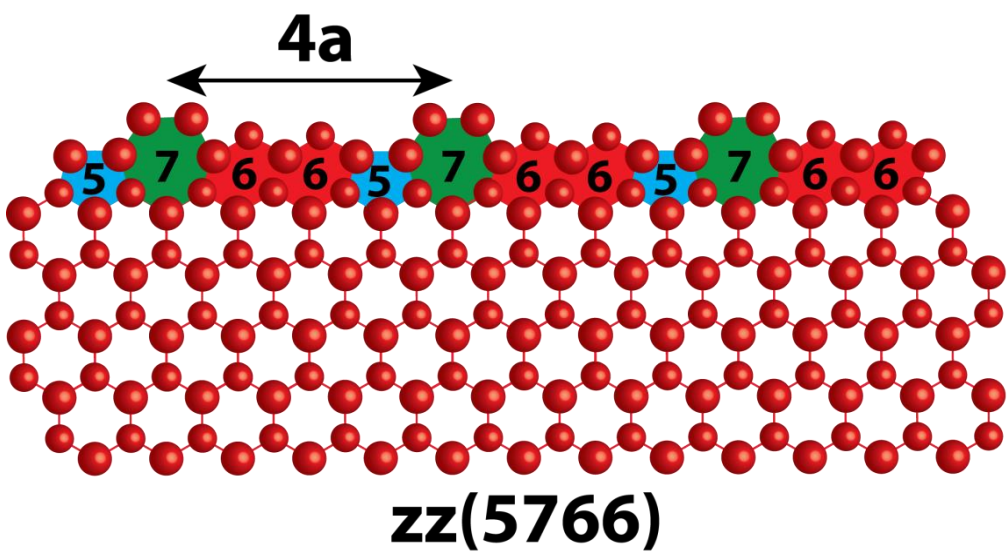

**Figure 4**

Schematic models of an unreconstructed zigzag edge (top panel), a zz(57) edge (middle panel) and a zz(5766) edge (bottom panel).